# Interaction des photons UV avec le silicium massif et nanocristallin

L. Patrone, I. Ozerov, M.L. Sentis[1] et W. Marine

Groupement Interdisciplinaire Ablation Laser et Applications, GPEC, UMR 6631 du CNRS, Faculté des Sciences de Luminy, Case 901, 13288 Marseille cedex 09, France
[1] Groupement Interdisciplinaire Ablation Laser et Applications, LP3, FRE 2165 du CNRS, Faculté des Sciences de Luminy, Case 901, 13288 Marseille cedex 09, France

**Résumé.** L'étude de l'interaction entre photons UV et le silicium massif et nanocristallin par spectrométrie de masse à temps de vol permet de distinguer deux populations de monomères ioniques $Si^+$. Une population rapide est éjectée dès les faibles fluences laser, dont l'origine est attribuée aux répulsions Coulombiennes entre les charges de surface créées par l'émission de photo-électrons. Alors que la fluence d'irradiation est augmentée, une population plus lente apparaît issue de phénomènes thermiques initiés dans le matériau irradié. La mise en évidence d'une population ionique non thermique qui constitue une part importante des ions présents dans la plume laser est un résultat original de cette étude.

## 1. INTRODUCTION

Parmi les nombreuses applications des lasers UV, leur utilisation dans le cadre de l'ablation laser a permis d'obtenir de très bons résultats concernant la préparation de nano-agrégats de Si [1]. La maîtrise des paramètres assurant le contrôle essentiel de la taille des nano-agrégats passe nécessairement par l'étude des espèces produites lors de l'irradiation de la cible de silicium massif. La forte dispersion des résultats publiés sur le sujet [2-6] montre la nécessité de poursuivre une telle étude. D'autre part, nous observons que les agrégats condensés dans la plume laser reviennent majoritairement se déposer sur la cible [1]. L'interaction entre photons UV et nano-agrégats déposés sur la zone irradiée intervient donc également au cours de la préparation des nano-agrégats. Une telle étude doit aussi permettre de mieux comprendre les mécanismes peu connus d'interaction entre les photons UV et les systèmes quantiques.

## 2. DISPOSITIF EXPERIMENTAL

La surface (111) du Si est préalablement désoxydée et passivée avec hydrogène par voie chimique. Les nano-agrégats de silicium étudiés (~ 1,5 nm de diamètre) sont préparés par ablation laser de la cible de silicium massif sous gaz de couverture inerte (0,76 Torr d'Ar). La désorption est réalisée sous vide ($10^{-8}$ Torr) avec un laser excimère ArF* impulsionnel ($\lambda$ = 193 nm, $\tau_p$ = 15 ns), sous un angle de 45° sur une aire de 0,5 mm² avec la partie la plus homogène du faisceau sélectionnée par des diaphragmes. La distribution temporelle des espèces se déplaçant le long de l'axe d'expansion de la plume est mesurée à l'aide d'un spectromètre de masse à temps de vol de type réflectron couplé à la chambre à vide. Le signal de temps de vol est alors l'intégrale de la surface du pic de la masse correspondante et de ses isotopes.

## 3. SILICIUM MASSIF

La désorption de monomères ioniques $Si^+$ est observée à partir d'un seuil de fluence d'irradiation de l'ordre de 0,2 J/cm² largement inférieur au seuil de fusion de ~ 0,4 J/cm² [7] et au seuil de détection des monomères neutres de ~ 0,8 J/cm². Cela indique un mécanisme d'éjection différent pour les ions et pour les particules neutres. La Fig. 1a montre les spectres de temps de vol des ions $Si^+$ mesurés pour différentes fluences variées entre 0,26 J/cm² et 1,02 J/cm². A faible fluence d'irradiation, une seule population d'ions Si$^+$ est éjectée avec une énergie cinétique de 4,8-5,8 eV correspondant au maximum du temps de vol à l'instant t = 20-22 µs. A partir de 0,42 J/cm², apparaît une population supplémentaire beaucoup moins énergétique (~ 3 eV, t ~ 28 µs) et possédant une distribution de vitesse plus large. Celle–ci se détache peu à peu de la première population qui est accélérée alors que la fluence est augmentée.

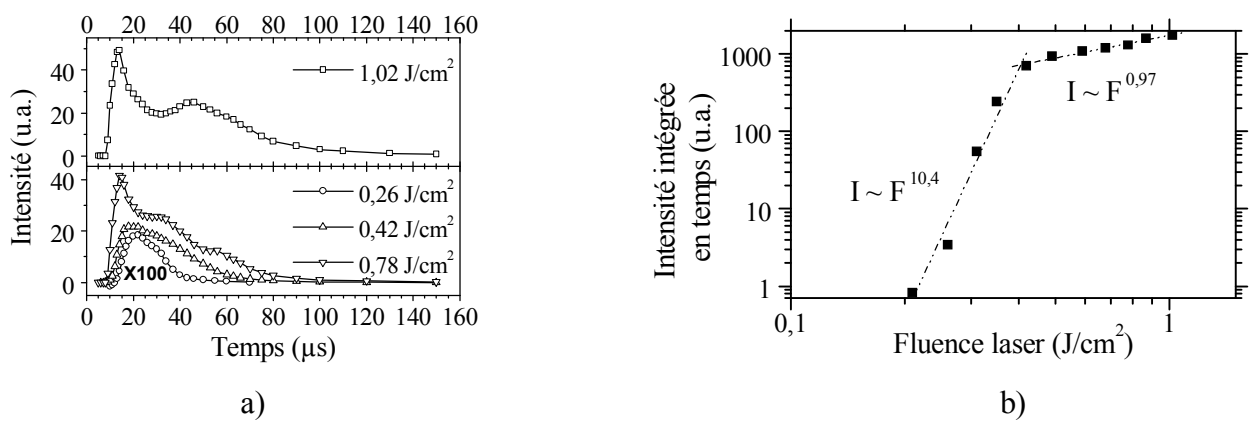

**Figure 1. a)** Temps de vol de l'ion Si$^+$ obtenu par désorption laser à partir de la surface du Si (111) irradié avec différentes fluences laser. L'intensité de la courbe mesurée à 0,26 J/cm² a été multipliée par 100. **b)** Variation de la quantité totale des deux populations d'ions Si$^+$ produits en fonction de la fluence laser d'irradiation. Distance cible – spectromètre = 126 mm.

La population rapide se caractérise par une distribution temporelle étroite de largeur à mi-hauteur à peu près constante avec la fluence laser (~ 15 µs). D'autre part, à l'instar d'autres études [2], nous avons mis en évidence sa forte directivité suivant la normale à la cible au moyen de l'inclinaison de la cible par rapport à l'axe du spectromètre de masse. Ces caractéristiques sont représentatives d'une éjection à partir d'effets photo-induits en surface [8], qui s'opèrent préférentiellement au niveau des défauts. Dans ces conditions, les ions rapides détectés sont issus d'un mécanisme d'éjection lié aux caractéristiques des photons incidents et/ou des propriétés de la surface irradiée. L'apparition de la population lente coïncide avec l'atteinte de la fluence du seuil de fusion du silicium (0,4 J/cm²). Son intensité croît avec la fluence laser jusqu'à devenir prépondérante au cours du régime d'ablation (0,8 J/cm²). Elle est donc liée à des phénomènes thermiques au niveau de la surface. La diminution de son énergie cinétique observée avec la fluence laser alors que la population rapide est fortement accélérée s'explique par la répulsion entre les deux populations ioniques.

L'analyse de l'évolution de la production ionique en fonction de la fluence laser représentée sur la Fig. 1b avec des échelles logarithmiques permet de distinguer deux régimes. Au-dessous de la fluence du seuil de fusion, la quantité d'ions produits augmente très rapidement suivant une puissance ~10 de la fluence laser montrant un processus d'éjection fortement non linéaire. A partir de la fusion, les ions sont majoritairement d'origine thermique. La quantité d'ions éjectés dépend alors peu de la fluence laser avec une variation quasi linéaire (puissance 0,97).

L'irradiation à faible fluence avec les photons utilisés d'énergie supérieure au travail de sortie du silicium (4,8 eV) entraîne l'émission de photo-électrons au niveau de la surface. Nous l'avons confirmé par la mesure d'une charge positive dynamique sur la cible de silicium à partir de fluences laser d'irradiation très faibles (~ 0,001 J/cm²). Le seuil d'éjection de photo-électrons que l'on observe est du même ordre que celui rapporté dans la littérature sur le silicium pour des photons d'énergie 6,4 eV [2]. Les charges positives ainsi créées conduisent à des répulsions Coulombiennes amenant à la brisure des liaisons des atomes de surface chargés, puis à leur éjection avec une énergie cinétique égale à la différence entre l'énergie Coulombienne répulsive et la somme des énergies de leurs liaisons. Ce type d'éjection est alors nécessairement directif, suivant la force Coulombienne exercée sur l'atome.

L'éjection des ions non thermiques est à l'origine de pertes dans l'énergie qui est transmise au matériau et qui conduit à sa fusion puis à sa vaporisation. Il en résulte l'augmentation de l'énergie nécessaire à la fusion d'une couche de surface d'épaisseur donnée (seuil dynamique de fusion) avec la fluence d'irradiation comme on peut le voir sur la Fig. 2. L'atteinte de la fusion du Si est détectée par l'accroissement de la réflectivité associé à la formation d'une couche liquide d'épaisseur fixée [9].

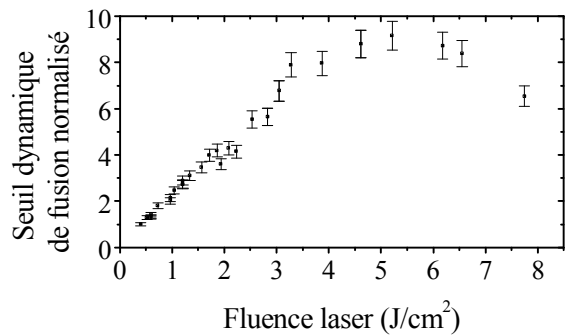

**Figure 2.** Variation de l'énergie nécessaire à la fusion de la surface du silicium (seuil dynamique de fusion) normalisée sur la valeur du seuil minimum (0,4 J/cm$^2$) en fonction de la fluence laser d'irradiation.

## 4. NANO-AGRÉGATS DE SILICIUM

On observe que l'irradiation des nano-agrégats de silicium conduit à l'éjection d'ions Si$^+$ jusqu'à des agrégats de 11 atomes Si$_{11}^+$ comme le montre le spectre de masse de la Fig. 3a. Les caractéristiques de la désorption ionique à partir des nano-agrégats irradiés montrent une forte analogie avec celles observées à partir du silicium massif, avec une éjection non thermique prononcée. La Fig. 3b représente l'intensité intégrée en temps $Y$ des spectres de temps de vol des ions Si$^+$ en fonction de la fluence laser $F$, tracée suivant des échelles logarithmiques. Les approximations en $F^\alpha$ sur les différentes parties de la courbe permettent d'évaluer le nombre $\alpha$ de photons simultanément engagés dans le régime de désorption. On distingue quatre régimes d'éjection ionique en fonction de la fluence laser.

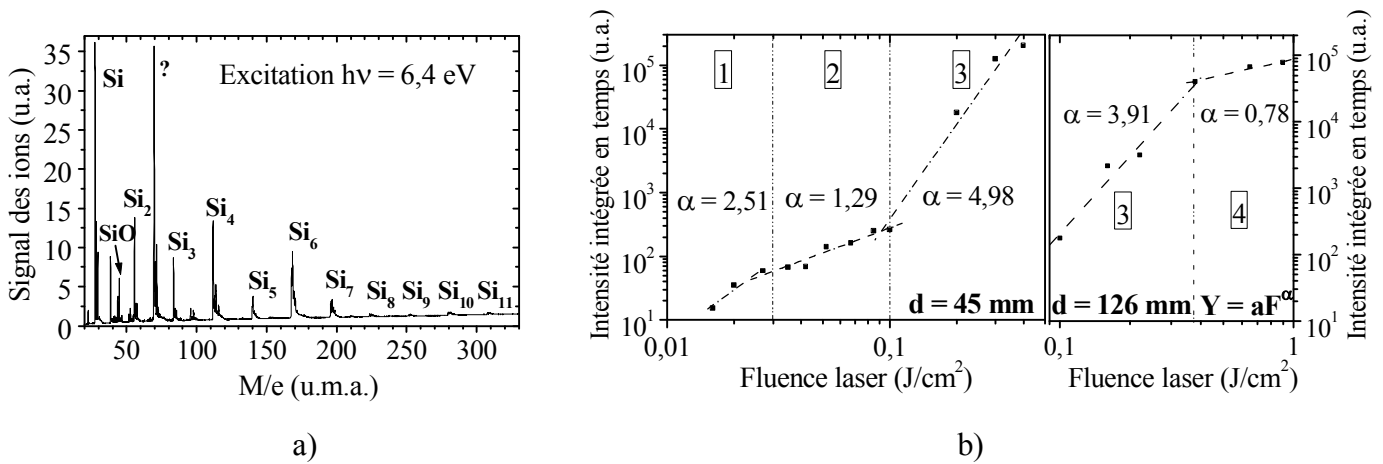

a)  b)

**Figure 3. a)** Spectre de masse typique des espèces éjectées au cours de l'irradiation des nano-agrégats de Si déposés. Fluence d'irradiation : 0,5 J/cm$^2$. **b)** Variation de l'intensité $Y$ des ions Si$^+$ désorbés à partir des nano-agrégats en fonction de la fluence laser $F$ (échelles logarithmiques). Deux distances cible – spectromètre de masse ont été utilisées: d = 45 mm pour les plus faibles fluences (0,02-0,4 J/cm$^2$), d = 126 mm pour les plus grandes fluences (0,1-0,9 J/cm$^2$).

L'éclairement des nano-agrégats de Si par les photons conduit à une excitation localisée des électrons pouvant entraîner l'éjection de photo-électrons grâce aux photons très énergétiques utilisés ($h\nu$ = 6,4 eV). Les charges positives ainsi créées à partir de la fluence seuil de ~ 0,016 J/cm$^2$ forment une charge d'espace en surface. L'addition de charges positives sur les nano-agrégats induit une forte déstabilisation des atomes et des petits agrégats de surface. Sous l'effet des forces Coulombiennes exercées sur ces derniers, le processus de désorption a lieu (Régime 1 : $F$ = 0,016-0,030 J/cm$^2$). La désorption commence d'abord par les atomes et les agrégats les plus faiblement liés. Au fur et à mesure que le nombre de charges s'accroît avec la fluence laser, les énergies potentielles de répulsion augmentent permettant l'éjection des espèces plus fortement liées. Cependant, la barrière de charge d'espace déjà formée rend plus difficile la création de nouvelles charges positives. Il en résulte une diminution du taux de production des ions $Si^+$ avec une dépendance en fonction de la fluence laser en puissance α plus faible (α = 1,29) (cf. Fig. 3b) et la stabilisation de l'énergie cinétique autour de 14-16 eV pour des fluences de 0,05-0,06 à 0,1 J/cm$^2$ (Régime 2 : $F$ = 0,03-0,10 J/cm$^2$). La chute de la production de petits agrégats correspond à la fin du processus de fragmentation et fait place à une décomposition massive en monomères $Si^+$ sous les effets de processus thermiques (fluences supérieures à 0,1 J/cm$^2$) au cours du régime 3 ($F$ = 0,1-0,4 J/cm$^2$). Ce dernier se caractérise sur la Fig. 3b par une augmentation rapide de la production ionique (α = 3,91 à la distance de 126 mm et α = 4,98 à la distance de 45 mm). L'éjection ionique est très vite dominée par les ions thermiques. La diminution de l'exposant α à plus grande distance de mesure peut être due à la divergence de la population thermique qui a été remarquée dans le cas de l'éjection à partir du silicium massif. La barrière de charges positives en surface empêche l'éjection de photo-électrons. L'absorption d'un photon entraîne alors une dissipation de son énergie à l'intérieur de l'agrégat sous forme d'énergie thermique. D'autre part, il faut également considérer le chauffage dû à l'absorption des porteurs libres piégés en grande concentration dans le nano-agrégat. L'énergie thermique est alors convertie en énergie cinétique transférée aux ions $Si^+$ produits par la décomposition des nano-agrégats. A partir d'une fluence d'environ 0,4 J/cm$^2$, les processus sont très majoritairement thermiques (Régime 4 : $F \geq 0,4$ J/cm$^2$). On observe l'éjection massive d'ions $Si^+$ thermiques. Leur nuage d'expansion d'abord lent subit une accélération hydrodynamique vers 0,65 J/cm$^2$. La production de ces ions montre une dépendance quasi linéaire en fonction de la fluence laser en $F^{0,78}$ caractéristique des processus d'émission thermique. En parallèle, la re-condensation massive des agrégats les plus lourds a lieu à partir des ions $Si^+$ et $Si_2^+$ produits en grande quantité.

## 4. CONCLUSION

L'étude de l'interaction entre les photons UV et le silicium massif et nanocristallin révèle deux populations distinctes de monomères ioniques. Dès les faibles fluences, on observe une population rapide dont l'éjection est interprétée en terme de répulsion Coulombienne à partir de la surface chargée par l'émission de photo-électrons. Cette population constitue une part importante des ions présents dans la plume laser. A partir du seuil de fusion apparaît une population plus lente et dont l'origine est attribuée à des phénomènes thermiques. L'éjection importante d'ions $Si^+$ et $Si_x^+$ (x=2-11) observée est bénéfique pour la synthèse de nano-agrégats de Si par ablation laser. L'augmentation de la population non thermique avec la fluence laser entraîne celle du seuil dynamique de fusion. Comme on l'a montré par ailleurs [10], cela se traduit par l'existence d'un régime de fluences laser pour lequel la température de la surface irradiée est stabilisée. Ces conditions sont favorables à la condensation de nano-agrégats de taille mieux contrôlée avec une dispersion réduite [1,10].